\documentclass{article}
\usepackage{spconf,amsmath,graphicx}
\usepackage{enumitem}
\usepackage{url}

\title{CORRGAN: SAMPLING REALISTIC FINANCIAL CORRELATION MATRICES USING GENERATIVE ADVERSARIAL NETWORKS}
\name{Gautier Marti}
\address{Shell Street Labs}

\begin{document}
%
\maketitle
\begin{abstract}
We propose a novel approach for sampling realistic financial correlation matrices. This approach is based on generative adversarial networks. Experiments demonstrate that generative adversarial networks are able to recover most of the known stylized facts about empirical correlation matrices estimated on asset returns. This is the first time such results are documented in the literature. Practical financial applications range from trading strategies enhancement to risk and portfolio stress testing. Such generative models can also help ground empirical finance deeper into science by allowing for falsifiability of statements and more objective comparison of empirical methods.
\end{abstract}
\section{Introduction}
\label{sec:intro}

In \cite{huttner2018portfolio}, we can read:
\begin{quote}
``To the best of our knowledge, there is no algorithm available for the generation of reasonably random
[financial] correlation matrices with the Perron-Frobenius property. 

[...]

Concerning the generation of [financial] correlation matrices whose MSTs [Minimum Spanning Trees] exhibit the scale-free property, to the best of our knowledge there is no algorithm available, and due to the generating mechanism of the MST we expect the task of finding such correlation matrices to be highly complex."
\end{quote}

In this paper, we propose a novel approach to solve the problem of generating realistic financial correlation matrices.
Using Generative Adversarial Networks (GANs) to sample realistic financial correlation matrices has never been documented, to the best of our knowledge, despite the importance of the problem. Simulating financial data, and correlation matrices in particular, have many applications: Testing robustness of trading strategies, stress testing portfolios. Another major application could be the objective comparison of empirical methods (combination of signals and strategies, statistical filtering methods \cite{tumminello2007shrinkage}) which would otherwise be claimed superior based on a given arbitrary chosen sample. This endemic problem in empirical finance prevents the field to become a science in the Popperian terminology: one cannot easily contradict such results \cite{lopez2019tactical}.

Generating multivariate financial time series is more general and difficult than to focus on their correlations: Besides the dependence structure (relatively static in comparison), one has to correctly capture the univariate time series features (e.g. autocorrelation) and the distributional properties of the margins altogether.
In this work, we only focus on generating empirical correlation matrices, which may already be an approximation of the dependence structure between several financial assets (cf. copula theory \cite{nelsen2007introduction}).









Despite the importance of the problem, we can explain the lack of research (and results) as GANs, a recent class of generative modelling approaches (seminal paper in 2014 \cite{goodfellow2014generative}) which stemmed from the computer science community, are not yet part of the econometrician, risk and quant analysts toolbox.
This work can also be relevant for the signal processing community as robust estimation of large covariance matrices $\Sigma$, since a correlation matrix $C = {\rm diag}(\Sigma)^{-\frac{1}{2}} \Sigma~ {\rm diag}(\Sigma)^{-\frac{1}{2}}$, is a common problem \cite{balaji2014information,aubry2017geometric}.

\subsection*{Contributions}
\label{sec:contrib}

The contributions of this article are:
\begin{itemize}[noitemsep]
    \item sampling financial correlation matrices using GANs, and documenting results for the first time,
    \item showing that the samples generated look realistic, and verify the stylized facts known in the econophysics literature,
    \item using S\&P 500 stock returns which are widely available for reproducibility of the experiments.
\end{itemize}{}

We also showcase our results through a web application: CorrGAN.io (\url{www.corrgan.io}). Users are asked to determine whether a given correlation matrix was generated from the GAN model or estimated from real stock returns. We have obtained a balanced number of correct and wrong answers so far. This is consistent with random guessing.

\section{Related work}
\label{sec:relwork}

To the best of our knowlege, there are no previous attempt at generating realistic financial correlation matrices using GANs. No known model is able to capture, even approximately, all the known characteristics of financial correlation matrices \cite{huttner2018portfolio}. 
We briefly review in the following subsection typical applications of GANs, and we highlight the lack of published results concerning financial data. Then, we describe the stylized facts of financial correlation matrices which will be useful to evaluate the samples generated by the different GAN-based approaches tested.

\subsection{Generative Adversarial Networks}
\label{sec:gans}

Generative Adversarial Networks (GANs) were introduced in \cite{goodfellow2014generative}.
Two networks $G$ (the generative model) and $D$ (a discriminative model) are trained simultaneously: $G$ is trained to maximize the probability of $D$ making a mistake; $D$ is trained to estimate the probability that a sample comes from the training data rather than $G$. These models are notoriously complex to train and evaluate. Their greatest success so far has been to generate realistic pictures. There are few successful applications published outside natural images, e.g. \cite{wu2016learning,binkowski2019high}; results produced by GANs are not competitive in natural language generation, for example.


Related to finance, authors are aware of \cite{henry2019generative} which aims at simulating SABR (a stochastic volatility model) parameters, and \cite{koshiyama2019generative} generating univariate time series of asset returns using a conditional GAN. The paper totally discards the whole dependence structure, e.g. correlations, existing between the time series of many assets. It may not matter when focusing on time series strategies (actively trading a single asset through time), but is useless when considering cross-sectional strategies, or large portfolio and risk management. In other words, it doesn't model the multivariate joint distribution of the co-movements of many assets. 

Unlike for natural images which lend themselves well to a visual inspection, samples obtained from GANs are in general hard to evaluate. Researchers are for now limited to check a few statistics, for example degree distribution of the graph nodes in \cite{bojchevski2018netgan} or the partial auto-correlation function of the time series in \cite{koshiyama2019generative}. However, one needs to know which statistics are important to verify. Fortunately, financial correlation matrices have been extensively researched over the past two decades.




\subsection{Financial correlation matrices}
\label{sec:fincorr}

Financial correlation matrices have been extensively studied in econophysics, an empirical field applying statistical physics methods to economy and finance. Around 1999, Bouchaud \textit{et al.} \cite{laloux2000random} showed how Random Matrix Theory (RMT) can be used to better understand financial correlations, and they started a two-decade research long program developing and refining methods using tools from RMT to clean large empirical correlation matrices \cite{bun2017cleaning}. About the same time, Mantegna, another econophysicist, discovered the hierarchical structure of financial correlations \cite{mantegna1999hierarchical} whose seminal and influential work sparked a rich empirical research in financial networks and clustering. An extensive review of this literature can be found in \cite{marti2017review}.

This body of knowledge about financial correlation matrices can be summarized in a few stylized facts:

\begin{itemize}[noitemsep]
\item Distribution of pairwise correlations is significantly shifted to the positive,
\item Eigenvalues follow the Marchenko-Pastur distribution \cite{laloux2000random}, but for
    \begin{itemize}[noitemsep]
        \item a very large first eigenvalue (the market),
        \item a couple of other large eigenvalues (industries),
    \end{itemize}
\item Perron-Frobenius property (first eigenvector has positive entries), 
\item Hierarchical structure of correlations \cite{mantegna1999hierarchical},  
\item Scale-free property of the corresponding Minimum Spanning Tree (MST) \cite{caldarelli2004emergence}.
\end{itemize}

It is possible that some stylized facts are still to be discovered.
Exploring the latent space of GANs \cite{chen2016infogan} could help finding unknown properties of financial correlations; However, generative adversarial networks, alongside deep learning in general, are not yet part of the toolkit in empirical finance. This paper is meant to show that they are a relevant tool, and that the problem of sampling financial correlation matrices using GANs deserves further exploration.

\section{The space of correlation matrices}
\label{sec:elliptope}

Let $C \in \mathbf{R}^{n \times n}$ be a correlation matrix, that is $C = C^\top$, $\forall i \in \{1, \ldots, n\}$, $C_{ii} = 1$, $\forall x \in \mathbf{R}^{n}$, $x^\top C x \geq 0$.

Let the elliptope of dimension $n(n-1)/2$ be the set corresponding to the $n(n-1)/2$ coefficients of $n \times n$-correlation matrix upper triangular. More formally,
\begin{equation*}
\begin{aligned}
& \mathcal{E}_{\frac{n(n-1)}{2}} = \{  \\ 
& \left(C_{12}, \ldots, C_{1n}, C_{23}, \ldots, C_{2n}, \ldots, C_{(n-1)n}\right) \in \mathbf{R}^{\frac{n(n-1)}{2}} ~|~  \\
& C = C^\top, \forall i \in \{1, \ldots, n\}, C_{ii} = 1, \forall x \in \mathbf{R}^{n}, x^\top C x \geq 0 \}
\end{aligned}
\end{equation*}


A $n \times n$-correlation matrix can be viewed as a point in $\mathcal{E}_{\frac{n(n-1)}{2}}$.

\subsection{$3 \times 3$ case}
\label{sec:3x3}

To build intuition, let's first consider the $3 \times 3$ case. We can visually verify that a simple GAN is able to recover the whole space of empirical correlations.

\begin{figure}[htb]
\begin{minipage}[b]{.48\linewidth}
  \centering
  \centerline{\includegraphics[width=4.0cm]{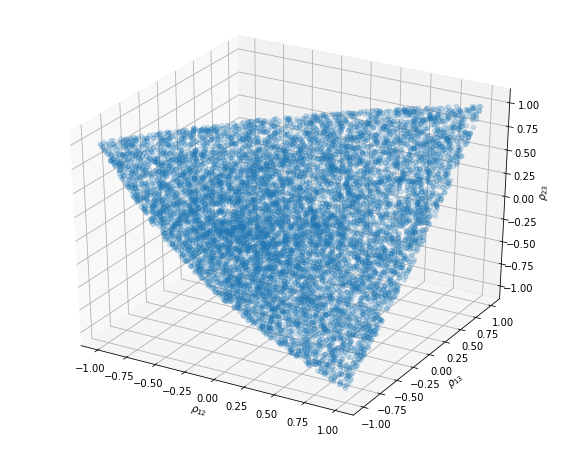}}
  \centerline{The $3$D elliptope}
\end{minipage}
\hfill
\begin{minipage}[b]{0.48\linewidth}
  \centering
  \centerline{\includegraphics[width=4.0cm]{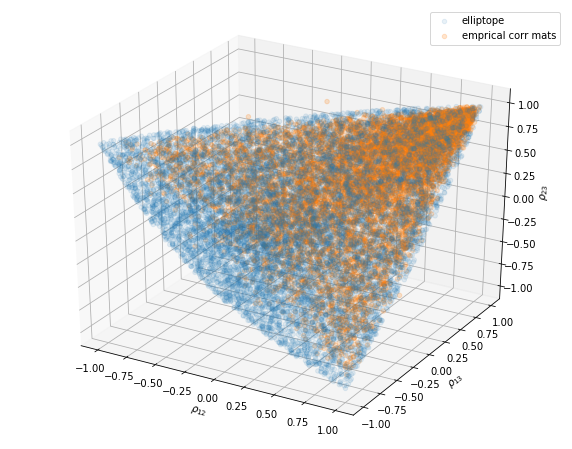}}
  \centerline{Empirical matrices (orange)}
\end{minipage}
\begin{minipage}[b]{1.0\linewidth}
  \centering
  \centerline{\includegraphics[width=8.5cm]{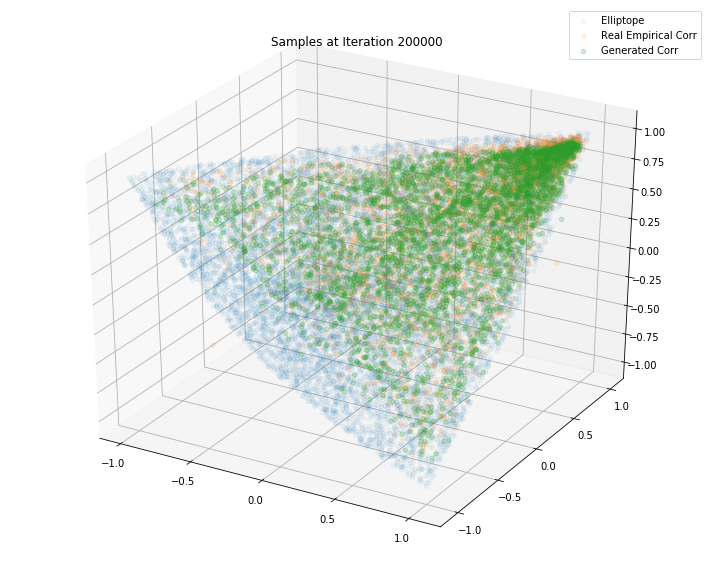}}
  \centerline{Sampling $3 \times 3$ correlations from a GAN (green points)}\medskip
\end{minipage}
\caption{We can visually inspect the results: A simple GAN is able to sample realistic $3 \times 3$ financial correlation matrices}
\label{fig:res}
\end{figure}

In Figure~\ref{fig:res}, 10,000 blue points are sampled uniformly (in the Lebesgue measure sense) from $\mathcal{E}_3$ using the onion method \cite{ghosh2003behavior}, where $\mathcal{E}_3$ is

$$\mathcal{E}_3 = \left\{ \left(\rho_{12}, \rho_{13}, \rho_{23} \right) \in \mathbf{R}^3 ~\middle|~ \begin{pmatrix}
1 & \rho_{12} & \rho_{13} \\ 
\rho_{12} & 1 & \rho_{23} \\ 
\rho_{13} & \rho_{23} & 1  
\end{pmatrix} \succeq 0 \right\}.$$

In orange, 10,000 3-by-3 matrices obtained by selecting randomly (without replacement) 3 stocks among the 500 possible in the S\&P 500, and then estimating the correlations between their daily returns on one year (252 business days). We can notice that the orange set (empirical correlations) is a strict subset of the blue set (whole space of valid correlation matrices) concentrated around average to high positive values. A simple GAN is able to recover this distribution (green points): It generates only valid correlation matrices, with a support matching closely the empirical ones, and a higher concentration around the average to high positive values.

\subsection{$n \times n$ case}
\label{sec:nxn}


In financial applications, $n$ ranges typically from a few dozens to a couple of hundreds, a few thousands in the most extreme cases. The large $n$ case is more difficult for many reasons, from statistical to computational. For our concerns, it is (i) harder to assess quality and coverage of samples generated and (ii) harder to train GANs as the standard neural networks are data inefficient on correlation matrices linked to their matrix equivalence property:  
When estimating a correlation matrix on a set of $n$ stock returns, the order of these stocks is arbitrary. There are $n!$ such possible orders, and therefore $n!$ different correlation matrices.
But they all essentially describe the same correlation structure.
We would like that the output of a neural network (here, the GAN discriminator (or critic) decision: \textit{fake} or \textit{real}) is invariant to permutations.



\begin{figure}[htb]
\begin{minipage}[b]{.48\linewidth}
  \centering
  \centerline{\includegraphics[width=\linewidth]{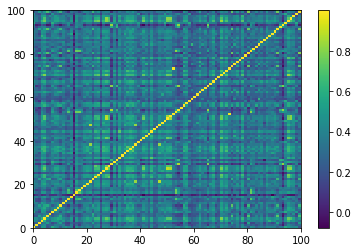}}
  \centerline{Arbitrary $C$}
\end{minipage}
\hfill
\begin{minipage}[b]{0.48\linewidth}
  \centering
  \centerline{\includegraphics[width=\linewidth]{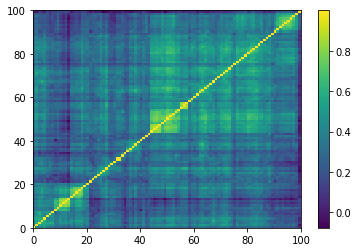}}
  \centerline{$R_{ij} = C_{\pi_H(i)\pi_H(j)}$}
\end{minipage}
\caption{Two equivalent correlation matrices. The one on the left has been obtained by estimation on returns of arbitrarily ordered stocks; The one on the right by applying $\pi_H$.}
\label{fig:repr}
\end{figure}

To solve this problem, we need to enforce permutation invariance either in the network (some early tentative in the literature \cite{zaheer2017deep}) or in the representation of the correlation matrix. The latter is the approach chosen in this work, namely we choose a representative for the equivalence class.
We propose to consider $R_{ij} = C_{\pi_H(i) \pi_H(j)}$, where $\pi_H$ is a permutation induced by a hierarchical clustering algorithm (inspired by one of the stylized facts, namely the hierarchical structure of financial correlations \cite{mantegna1999hierarchical})
We show in Figure~\ref{fig:repr} the result of applying $\pi_H$ to a given correlation matrix.

\section{Results and Evaluation}
\label{sec:eval}


We apply a deep convolutional generative adversarial network (DCGAN) \cite{radford2015unsupervised}, whose architecture is known to be able to learn a hierarchy of representations from object parts to scenes in natural images, on approximately 10,000 empirical correlation matrices estimated on S\&P 500 returns sorted using $\pi_H$.
Note that the matrices generated by the GAN models are not exactly correlation matrices: Their diagonal is not exactly equal to 1 (coefficients obtained are around 0.998); the matrices look visually symmetric but are not; Small negative eigenvalues can be found. We post process the results using an alternating projections method described in \cite{higham2002computing} to find the nearest correlation matrix with respect to the Frobenius norm.

\begin{figure}[htb]
    \begin{minipage}[b]{.48\linewidth}
      \centering
      \centerline{\includegraphics[width=\linewidth]{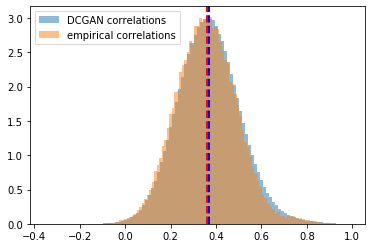}}
      \centerline{Distribution of correlations}
    \end{minipage}
    \hfill
    \begin{minipage}[b]{0.48\linewidth}
      \centering
      \centerline{\includegraphics[width=\linewidth]{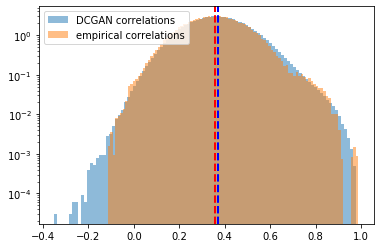}}
      \centerline{Log distribution}
    \end{minipage}
    \caption{The distributions of empirical and DCGAN-generated correlation coefficients match closely: They have approximately the same mean (0.36) and standard deviation (0.13). We can notice in the log-plot some discrepancies in the tails.}
    \label{fig:coeffs}
\end{figure}

\begin{figure}[htb]
    \begin{minipage}[b]{.48\linewidth}
      \centering
      \centerline{\includegraphics[width=\linewidth]{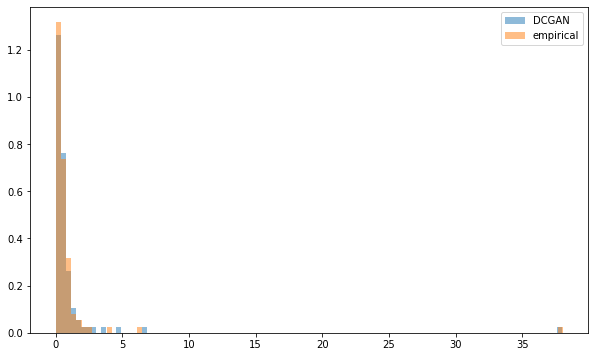}}
      \centerline{Distribution of eigenvalues}
    \end{minipage}
    \hfill
    \begin{minipage}[b]{0.48\linewidth}
      \centering
      \centerline{\includegraphics[width=\linewidth]{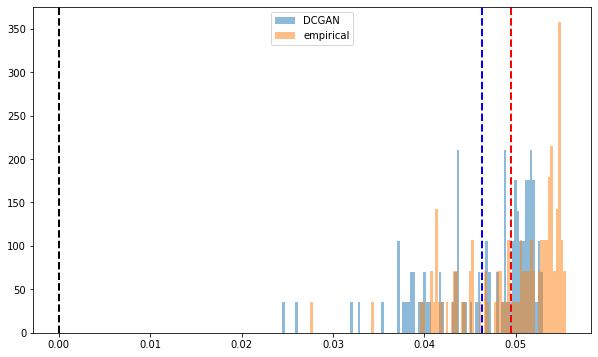}}
      \centerline{First eigenvector entries}
    \end{minipage}
    \caption{(Left) We can notice that the synthetic eigenvalues distribution share similar characteristics, i.e. a very large first eigenvalue, and a few ones outside the bulk of the distribution; (Right) All entries of the dominant eigenvector are positives.}
    \label{fig:eigs}
\end{figure}

\begin{figure}[htb]
\begin{minipage}[b]{.32\linewidth}
  \centering
  \centerline{\includegraphics[width=\linewidth]{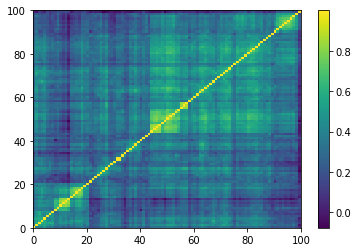}}
  \centerline{}
\end{minipage}
\hfill
\begin{minipage}[b]{.32\linewidth}
  \centering
  \centerline{\includegraphics[width=\linewidth]{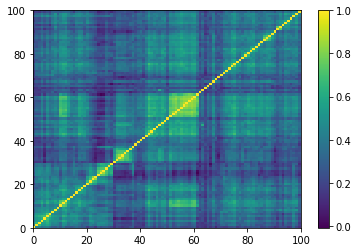}}
  \centerline{}
\end{minipage}
\hfill
\begin{minipage}[b]{0.32\linewidth}
  \centering
  \centerline{\includegraphics[width=\linewidth]{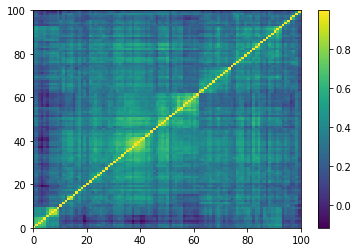}}
  \centerline{}
\end{minipage}
\begin{minipage}[b]{.32\linewidth}
  \centering
  \centerline{\includegraphics[width=\linewidth]{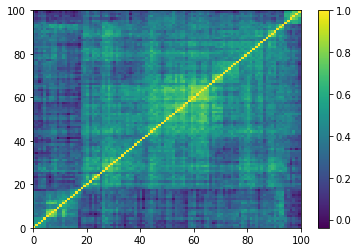}}
  \centerline{}
\end{minipage}
\hfill
\begin{minipage}[b]{.32\linewidth}
  \centering
  \centerline{\includegraphics[width=\linewidth]{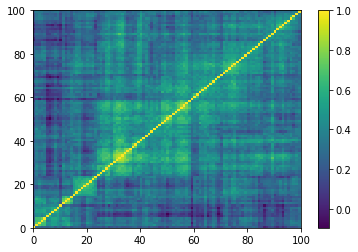}}
  \centerline{}
\end{minipage}
\hfill
\begin{minipage}[b]{0.32\linewidth}
  \centering
  \centerline{\includegraphics[width=\linewidth]{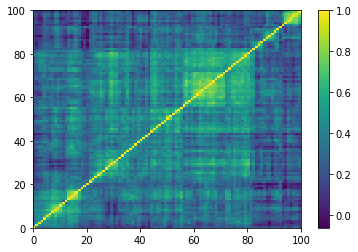}}
  \centerline{}
\end{minipage}

\caption{Top row: Three randomly selected empirical correlation matrices; Bottom row: Three DCGAN-generated correlation matrices. We can notice the existence of hierarchical clusters in both set of matrices.}
\label{fig:mats}
\end{figure}

\begin{figure}[htb]
    \includegraphics[width=\linewidth]{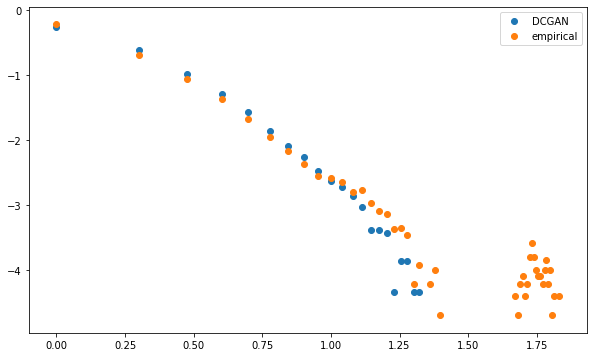}
    \caption{Log-log plot of the distribution of node degrees in the MST. The DCGAN captures well the distribution of degrees (seemingly following a power law) but for the tail: A very few nodes have very high degrees. Typically, General Electric is known to occupy a central position in the S\&P 500 MST.}
    \label{fig:powerlaw}
\end{figure}

Results obtained are evaluated using the stylized facts described in section~\ref{sec:fincorr}: Do we recover the main characteristics of financial correlation matrices? Essentially, yes. Tails of the distributions are not perfectly simulated though.
Comparison between empirical and synthetic samples are displayed in Figures~\ref{fig:coeffs},~\ref{fig:eigs},~\ref{fig:mats},~\ref{fig:powerlaw}.

Another experiment we did to assess the results was to ask people to determine whether a given matrix is \textit{real}, i.e. estimated from stocks returns, or \textit{fake}, i.e. generated from the GAN model. Concretely, we built a web application: CorrGAN.io (\url{www.corrgan.io}), which displays samples from both classes with equal probability, and where users can input their guess.
We have obtained a balanced number of correct and wrong answers so far: Samples generated by the GAN seem also realistic to the human eye.

\section{Discussion}
\label{sec:discussion}

We have proposed a novel approach using GANs to generate realistic financial correlation matrices.
The approach can be perfected, notably by spending more time and resources on experimental settings, but results showcased in this work are convincing. With this new tool, we can, for example, revisit the results described in \cite{huttner2018portfolio} quoted in our introduction, which compares portfolios based on graphs to Markowitz-optimal portfolios.

It would be interesting to explore the use of Topological Data Analysis to compare the empirical data manifold to the synthetic data manifold as proposed in \cite{khrulkov2018geometry}. In this paper, we verified that the generated samples are realistic, but do they span the whole subspace of realistic financial correlation matrices? We might only sample from a restricted part of the space, for example due to a mode collapse during the GAN training.

This work could be an important component in improving Monte Carlo backtesting \cite{lopez2019tactical}: Many paths can be sampled from a multivariate distribution parameterized by GAN-generated correlation matrices. Exploring conditional generation, for example conditioning on a market regime variable (risk-on or risk-off; quantitative easing or quantitative tightening; global crisis or not), could lead to new ways of stress testing portfolios.

Finally, investigating the latent space of these models could lead to a better understanding of financial correlations, and maybe the discovery of unknown stylized facts.

\vfill\pagebreak

\bibliographystyle{IEEEbib}
\bibliography{strings,refs}

\begin{thebibliography}{10}

\bibitem{huttner2018portfolio}
Amelie H{\"u}ttner, Jan-Frederik Mai, and Stefano Mineo,
\newblock ``Portfolio selection based on graphs: Does it align with
  markowitz-optimal portfolios?,''
\newblock {\em Dependence Modeling}, vol. 6, no. 1, pp. 63--87, 2018.

\bibitem{tumminello2007shrinkage}
Michele Tumminello, Fabrizio Lillo, and Rosario~Nunzio Mantegna,
\newblock ``Shrinkage and spectral filtering of correlation matrices: a
  comparison via the {K}ullback-{L}eibler distance,''
\newblock {\em arXiv preprint arXiv:0710.0576}, 2007.

\bibitem{lopez2019tactical}
Marcos Lopez~de Prado,
\newblock ``Tactical investment algorithms,''
\newblock {\em Available at SSRN 3459866}, 2019.

\bibitem{nelsen2007introduction}
Roger~B Nelsen,
\newblock {\em An introduction to copulas},
\newblock Springer Science \& Business Media, 2007.

\bibitem{goodfellow2014generative}
Ian Goodfellow, Jean Pouget-Abadie, Mehdi Mirza, Bing Xu, David Warde-Farley,
  Sherjil Ozair, Aaron Courville, and Yoshua Bengio,
\newblock ``Generative adversarial nets,''
\newblock in {\em Advances in neural information processing systems}, 2014, pp.
  2672--2680.

\bibitem{balaji2014information}
Bhashyam Balaji, Fred{\'e}ric Barbaresco, and Alexis Decurninge,
\newblock ``Information geometry and estimation of {T}oeplitz covariance
  matrices,''
\newblock in {\em 2014 International Radar Conference}. IEEE, 2014, pp. 1--4.

\bibitem{aubry2017geometric}
Augusto Aubry, Antonio De~Maio, and Luca Pallotta,
\newblock ``A geometric approach to covariance matrix estimation and its
  applications to radar problems,''
\newblock {\em IEEE Transactions on Signal Processing}, vol. 66, no. 4, pp.
  907--922, 2017.

\bibitem{wu2016learning}
Jiajun Wu, Chengkai Zhang, Tianfan Xue, Bill Freeman, and Josh Tenenbaum,
\newblock ``Learning a probabilistic latent space of object shapes via 3d
  generative-adversarial modeling,''
\newblock in {\em Advances in neural information processing systems}, 2016, pp.
  82--90.

\bibitem{binkowski2019high}
Miko{\l}aj Bi{\'n}kowski, Jeff Donahue, Sander Dieleman, Aidan Clark, Erich
  Elsen, Norman Casagrande, Luis~C Cobo, and Karen Simonyan,
\newblock ``High fidelity speech synthesis with adversarial networks,''
\newblock {\em arXiv preprint arXiv:1909.11646}, 2019.

\bibitem{henry2019generative}
Pierre Henry-Labordere,
\newblock ``Generative models for financial data,''
\newblock {\em Available at SSRN 3408007}, 2019.

\bibitem{koshiyama2019generative}
Adriano Koshiyama, Nick Firoozye, and Philip Treleaven,
\newblock ``Generative adversarial networks for financial trading strategies
  fine-tuning and combination,''
\newblock {\em arXiv preprint arXiv:1901.01751}, 2019.

\bibitem{bojchevski2018netgan}
Aleksandar Bojchevski, Oleksandr Shchur, Daniel Z{\"u}gner, and Stephan
  G{\"u}nnemann,
\newblock ``{NetGAN}: Generating graphs via random walks,''
\newblock {\em arXiv preprint arXiv:1803.00816}, 2018.

\bibitem{laloux2000random}
Laurent Laloux, Pierre Cizeau, Marc Potters, and Jean-Philippe Bouchaud,
\newblock ``Random matrix theory and financial correlations,''
\newblock {\em International Journal of Theoretical and Applied Finance}, vol.
  3, no. 03, pp. 391--397, 2000.

\bibitem{bun2017cleaning}
Jo{\"e}l Bun, Jean-Philippe Bouchaud, and Marc Potters,
\newblock ``Cleaning large correlation matrices: tools from random matrix
  theory,''
\newblock {\em Physics Reports}, vol. 666, pp. 1--109, 2017.

\bibitem{mantegna1999hierarchical}
Rosario~N Mantegna,
\newblock ``Hierarchical structure in financial markets,''
\newblock {\em The European Physical Journal B-Condensed Matter and Complex
  Systems}, vol. 11, no. 1, pp. 193--197, 1999.

\bibitem{marti2017review}
Gautier Marti, Frank Nielsen, Miko{\l}aj Bi{\'n}kowski, and Philippe Donnat,
\newblock ``A review of two decades of correlations, hierarchies, networks and
  clustering in financial markets,''
\newblock {\em arXiv preprint arXiv:1703.00485}, 2017.

\bibitem{caldarelli2004emergence}
Guido Caldarelli, Stefano Battiston, Diego Garlaschelli, and Michele Catanzaro,
\newblock ``Emergence of complexity in financial networks,''
\newblock in {\em Complex Networks}, pp. 399--423. Springer, 2004.

\bibitem{chen2016infogan}
Xi~Chen, Yan Duan, Rein Houthooft, John Schulman, Ilya Sutskever, and Pieter
  Abbeel,
\newblock ``{InfoGAN}: Interpretable representation learning by information
  maximizing generative adversarial nets,''
\newblock in {\em Advances in neural information processing systems}, 2016, pp.
  2172--2180.

\bibitem{ghosh2003behavior}
Soumyadip Ghosh and Shane~G Henderson,
\newblock ``Behavior of the {NORTA} method for correlated random vector
  generation as the dimension increases,''
\newblock {\em ACM Transactions on Modeling and Computer Simulation (TOMACS)},
  vol. 13, no. 3, pp. 276--294, 2003.

\bibitem{zaheer2017deep}
Manzil Zaheer, Satwik Kottur, Siamak Ravanbakhsh, Barnabas Poczos, Ruslan~R
  Salakhutdinov, and Alexander~J Smola,
\newblock ``Deep sets,''
\newblock in {\em Advances in neural information processing systems}, 2017, pp.
  3391--3401.

\bibitem{radford2015unsupervised}
Alec Radford, Luke Metz, and Soumith Chintala,
\newblock ``Unsupervised representation learning with deep convolutional
  generative adversarial networks,''
\newblock {\em arXiv preprint arXiv:1511.06434}, 2015.

\bibitem{higham2002computing}
Nicholas~J Higham,
\newblock ``Computing the nearest correlation matrix - a problem from
  finance,''
\newblock {\em IMA journal of Numerical Analysis}, vol. 22, no. 3, pp.
  329--343, 2002.

\bibitem{khrulkov2018geometry}
Valentin Khrulkov and Ivan Oseledets,
\newblock ``Geometry score: A method for comparing generative adversarial
  networks,''
\newblock {\em arXiv preprint arXiv:1802.02664}, 2018.

\end{thebibliography}

\end{document}